\documentclass[twocolumn]{aastex63}
\usepackage{soul}

\begin{document}

\title{Photometric variability as a proxy for magnetic activity and its dependence on metallicity}
\author{Victor See$^{1}$, Julia Roquette$^{1}$, Louis Amard$^{1}$, Sean P. Matt}
\affil{University of Exeter, Deparment of Physics \& Astronomy, Stocker Road, Devon, Exeter, EX4 4QL, UK}
\email{*w.see@exeter.ac.uk}

\begin{abstract}
Understanding how the magnetic activity of low-mass stars depends on their fundamental parameters is an important goal of stellar astrophysics. Previous studies show that activity levels are largely determined by the stellar Rossby number which is defined as the rotation period divided by the convective turnover time. However, we currently have little information on the role that chemical composition plays. In this work, we investigate how metallicity affects magnetic activity using photometric variability as an activity proxy. Similarly to other proxies, we demonstrate that the amplitude of photometric variability is well parameterised by the Rossby number, although in a more complex way. We also show that variability amplitude and metallicity are generally positively correlated. This trend can be understood in terms of the effect that metallicity has on stellar structure and, hence, the convective turnover time (or, equivalently, the Rossby number). Lastly, we demonstrate that the metallicity dependence of photometric variability results in a rotation period detection bias whereby the periods of metal-rich stars are more easily recovered for stars of a given mass.
\end{abstract}


\section{Introduction} 
\label{sec:Intro}
Attempting to understand how magnetic activity scales with fundamental stellar parameters in low-mass stars ($M_\star \lesssim 1.3M_\odot$) is an ongoing task within stellar astrophysics. It is well known that many forms of magnetic activity can be parameterised as a function of the Rossby number, Ro, including X-ray emission \citep{Pizzolato2003,Wright2018}, chromospheric Ca II H \& K line emission \citep{Noyes1984,Saar1999,Mamajek2008}, UV emission \citep{Stelzer2016}, H$\alpha$ emission \citep{Newton2017} as well as the strength of magnetic fields themselves \citep{Reiners2009,Vidotto2014ZDI,See2019ZB,Kochukhov2020}. This dimensionless number, defined as the rotation period divided by the convective turnover time, encapsulates the interplay between rotation and convection that is thought to be responsible for driving dynamo action \citep[e.g.][]{Brun2017}. In general, stars with smaller Rossby numbers are more magnetically active until a saturation level is reached for stars with the smallest Rossby numbers. However, it is worth noting that the Rossby number is difficult to estimate. Since the convective turnover time is not a directly observable quantity, any estimates of it, and therefore the Rossby number, will be model dependent.

While many works have studied how magnetic activity depends on the Rossby number, relatively few have studied the impact of metallicity (primarily due to lack of metallicity data for large samples of stars). From a theoretical perspective, one might expect that activity would be affected by metallicity through its impact on a star's internal structure. More metal-rich gas is expected to be more opaque, resulting in a deeper stellar convection zone and a longer convective turnover time. At fixed mass and rotation period, more metal-rich stars would therefore have a smaller Rossby number \citep{Karoff2018,Amard2019}. Since activity seems to largely depend on just the Rossby number, it would be reasonable to suppose that more metal-rich stars are more magnetically active. This is the central hypothesis that we will explore in this paper.

In order to conduct the type of study we present here, we require a magnetic activity proxy that can be easily determined for many stars. Traditional proxies such as X-ray or chromospheric emission are comparatively hard to determine for large stellar samples and so we use a more indirect tracer of magnetic activity, the photometric variability amplitude, $R_{\rm per}$. This is a quantity that is determined from a star's light curve which means it can be estimated for large numbers of stars due to photometric surveys like the Kepler mission. It is typically defined as the difference between the 5th and 95th percentiles of a star's light curve after it has been normalised by its median flux, although slight variations on this definition are also sometimes employed. It was first introduced by \citet{Basri2010} and subsequent works have investigated how the variability amplitude varies with fundamental stellar parameters \citep[e.g.][]{Basri2011,Reinhold2013,McQuillan2014,Reinhold2020}. Since the variations in light curves are induced by magnetic features such as dark spots or bright faculae, more magnetically active stars generally have larger variability amplitudes. Previous work has shown that the variability amplitude is generally larger for more rapidly rotating stars \citep{Reinhold2013,McQuillan2014}, as one would expect of an activity proxy, and also that it correlates with other activity proxies \citep{Radick1998,Karoff2016}. Additionally, stellar activity has also been shown to correlate well with $S_{\rm ph}$, which is another measure of photometric variability that is similar to $R_{\rm per}$ \citep{Salabert2016}. Together, these works demonstrate that the photometric variability is a good proxy for magnetic activity.

There have already been investigations into the effect of metallicity on photometric variability, and therefore, magnetic activity. For example, \citet{Karoff2018} compared the variability amplitudes of the Sun and a more metal-rich solar analogue, HD 173701. While this was a comparison of only two stars, these authors found that the more metal-rich star has a larger variability which agrees with the theoretical expectation outlined above. Additionally, in their supplementary materials, \citet{Reinhold2020} fit a multivariate linear regression to a sample of stars similar to the one we use in this work. This regression expressed the variability amplitude in terms of effective temperature, rotation period and metallicity. Again, in line with the theoretical expectation, the authors found that the variability amplitude had a positive dependence on the metallicity term in their fit although we note that the terms in this fit were not independent of each other. Specifically, effective temperature also depends on metallicity. This makes it hard to determine the exact dependence that variability amplitude has on metallicity alone from the multivariate fit. Lastly, there have been a number of studies investigating the brightness contrasts associated with different magnetic features, e.g. spots and faculae, and how they affect the overall brightness variability on stars on various time-scales \citep{Shapiro2014,Shaprio2016,Witzke2018,Witzke2020}. These works indicate that the brightness contrasts are a complicated function of many factors including, but not limited to, the fraction of stellar surfaces covered by magnetic features, the inclination angle and, indeed, the metallicity of the star which can affect the strength of Fraunhofer lines that contribute to a star's brightness.

In this work, we will empirically investigate the relationship between metallicity and variability amplitude using a sample of over 3000 stars. The rest of this paper is structured as follows. In section \ref{sec:Sample}, we present the sample of stars that we use in this work. The activity trends seen in our sample are discussed in section \ref{sec:Trends}. In section \ref{sec:RotRecovery}, we discuss the implications of these activity trends for measuring rotation periods from light curves. In section \ref{sec:Disc}, we discuss several aspects of this work in more depth as well as its implications for other studies. Lastly, a summary is presented in section \ref{sec:Summary}.

\section{Sample}
\label{sec:Sample}
In this section, we present the samples of stars we use throughout the rest of this work. Our samples are based on those of \citet{McQuillan2014}. Using an autocorrelation method, these authors attempted to measure the rotation periods of $\sim$133,000 stars from Kepler light curves. Of these stars, they successfully obtained a rotation period for $\sim$34,000 stars (which we denote as the periodic sample) while no period was recovered for the remaining $\sim$99,000 stars (which we denote as the non-periodic sample). In addition to rotation periods, \citet{McQuillan2014} also compiled variability amplitudes for the stars in the periodic sample. The specific method used to calculate variability amplitude by these authors is as follows. First, each light curve is divided into segments equal in length to the rotation period of the star. The difference between the 5th and 95th percentile of the normalised flux is then calculated for each segment. The reported variability amplitude,  $R_{\rm per}$, is the median value of these differences across every segment. This method ensures that $R_{\rm per}$ is a measure of variability over rotational time-scales rather than, e.g. activity cycle time-scales.

Next, we cross-matched the periodic and non-periodic samples with the Gaia-Kepler catalogue \citep{GaiaKepler} to obtain distances and photometry for our stars. This catalogue is, itself, a cross-match between the Gaia DR2 source catalogue and the Kepler DR25 catalog and contains over 200,000 Kepler field sources. Following the recommendations of this catalogue, we exclude all duplicate sources and select only those with high quality Gaia DR2 data, with parallax error lower than 0.1 mas and photometric error lower than 1 percent in every photometric band. The Gaia-Kepler catalogue also includes an improved distance prescription from \citet{BailerJones2018}. We only use sources that do not have a bimodal distance solution and have a well-constrained distance in Gaia DR2. Using these distances, $d$, we convert the Gaia apparent magnitudes, $G$, to absolute magnitudes, $M_G=G-5\log_{10}(\frac{d}{10})$. After this process, the periodic and non-periodic samples contain 28,508 and 86,824 stars respectively. 

\begin{figure}
	\begin{center}
	\includegraphics[trim=0cm 0.8cm 0cm 0.8cm,width=\columnwidth]{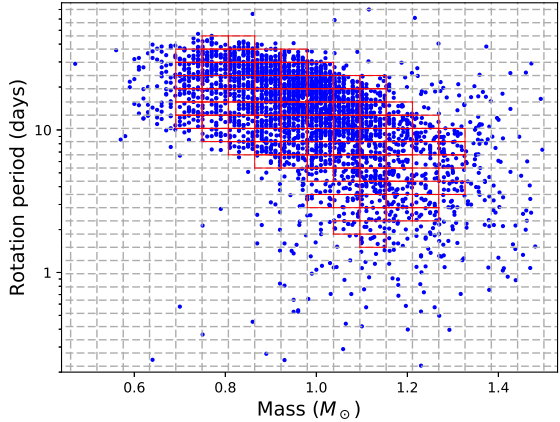}
	\end{center}
	\caption{The sample of stars used for this study in period-mass space. The dashed gray lines indicate the period and mass bins used throughout this work. Bins that contain at least 10 stars are outlined in red.}
	\label{fig:PerMass}
\end{figure}

For metallicities, [Fe/H], and effective temperatures, $T_{\rm eff}$, we adopt the values from LAMOST DR5 \citep{Luo2019}. This catalogue provides stellar parameters for millions of stars based on mid-resolution (R$\sim$1800) spectra and includes $\sim$190,000 sources in the Kepler Field. We only select sources that have [Fe/H] values with a reported precision better than 0.1 dex, which eliminates most [Fe/H] measurements obtained from spectra with a signal-to-noise worse than $\sim$50. After merging with LAMOST DR5, the periodic and non-periodic samples are reduced to 6033 stars and 16941 stars respectively. Both the metallicity and effective temperature are used along with the Gaia DR2 absolute photometry to estimate masses for our samples using the grid of stellar evolution models of \citet[][STAREVOL]{Amard2019} and an adapted maximum-likelihood interpolation tool \citep{Valle2014}. Thanks to the relatively high precision of the input observables ($T_{\rm eff}$, [Fe/H] and magnitudes), these mass estimates have only a 5\% error at most. Note that this error does not account for systematic uncertainties associated with the physics of the stellar evolution code.

As well as masses, the stellar structure grid of \citet{Amard2019} also gives the convective turnover time for each star. For this work, we use the mixing length theory description \citep{Charbonnel2017} where turnover times are defined as the ratio of the mixing length over the convective velocity taken at one pressure scale height above the base of the convective envelope. The mixing length parameter, $\alpha$, is chosen to be 1.973. In this formalism the solar convective turnover time is 15 days corresponding to a solar Rossby number of $\sim$1.8 days. Since turnover time estimates are model dependent, our convective turnover times will differ slightly to other estimates that make different assumptions (see appendix \ref{sec:Appendix} for further discussion on our choice of turnover time). However, we have chosen to use the convective turnover times from the structure grid of \citet{Amard2019} rather than other common prescriptions so that our turnover times are consistent with our mass estimates and because it allows us to explicitly account for the metallicity of the stars in our sample.

\begin{figure}
	\begin{center}
	\includegraphics[trim=0cm 1.0cm 0cm 0.4cm,width=\columnwidth]{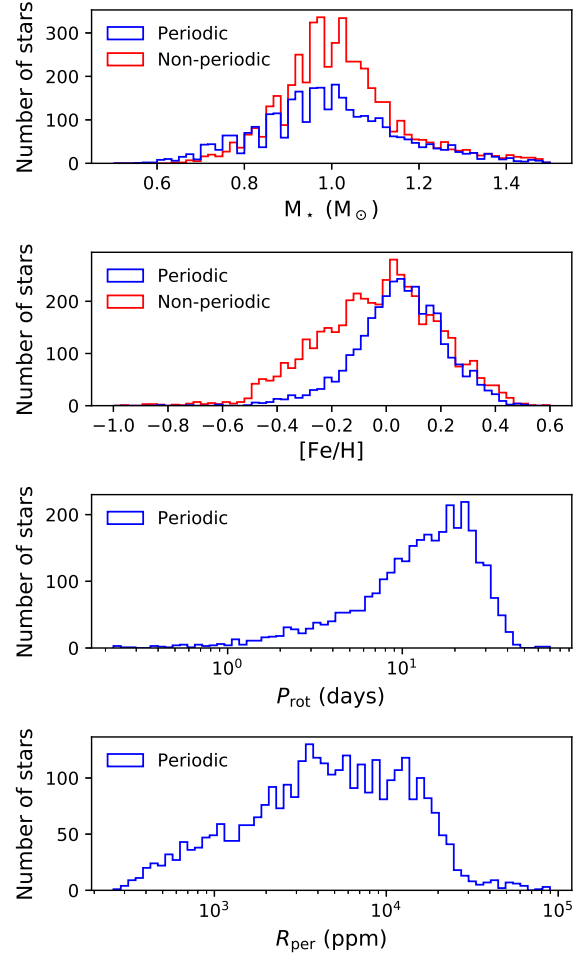}
	\end{center}
	\caption{Mass, $M_\star$, metallicity, [Fe/H], rotation period, $P_{\rm rot}$, and variability amplitude, $R_{\rm per}$, histograms for the periodic (blue) and non-periodic (red; where available) samples.}
	\label{fig:Hists}
\end{figure}

Finally, we eliminate binaries and evolved objects following the method we previously used in \citet{Amard2020}. First, we remove all stars further than 1 kpc from our sample. This cut means that most of the sample is only affected by low extinction ($A_V\lesssim0.213$ mag) and also naturally eliminates many background giants and sub-giants. Secondly, we eliminate possible nearly equal-mass binaries that are typically located $2.5\log(2)\simeq 0.753$ mag above the main sequence using a set of metallicity dependent cuts. We refer the reader to our previous work for full details of this process \citep{Amard2020}. After these cuts are made, our final periodic and non-periodic samples contain 3232 stars and 4746 stars respectively. The stellar parameters of these samples (mass, metallicity and convective turnover time for both samples as well as rotation period and variability amplitude for the periodic sample) can be found online at CDS. Figure \ref{fig:PerMass} shows the periodic sample in rotation period-mass space while fig. \ref{fig:Hists} shows mass, metallicity, rotation period and variability amplitude histograms for both the periodic and non-periodic samples. 

\section{Activity trends}
\label{sec:Trends}

\subsection{Trends with Rossby number}
\label{subsec:RossbyTrends}
In general, the magnetic activity of slowly rotating stars in the so-called unsaturated regime has an inverse power-law dependence on the Rossby number. However, at small Rossby numbers, the magnetic activity of rapidly rotating stars in the so-called saturated regime plateaus to a constant value \citep[see][and references therein for a discussion of these regimes in the context of X-ray emission]{Wright2011}. The transition between these two regimes occurs at a Rossby number of Ro$\sim$0.1 although this value can vary slightly depending on the activity indicator in question and the method used to estimate the convective turnover time.

\begin{figure}
	\begin{center}
	\includegraphics[trim=0cm 0.8cm 0.8cm 0.8cm,width=\columnwidth]{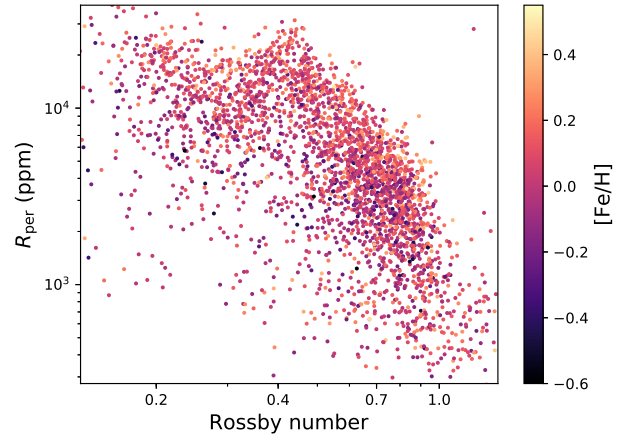}
	\end{center}
	\caption{Photometric variability amplitude against Rossby number coloured by metallicity.}
	\label{fig:ActRoss}
\end{figure}

In fig \ref{fig:ActRoss}, we show the variability amplitude as a function of Rossby number for our sample of stars. The variability amplitudes of the $\rm Ro\gtrsim 0.4$ stars scale inversely with Rossby number. This is broadly consistent with the results of past studies on variability amplitude that show that the variability is larger for more rapidly rotating stars \citep{Reinhold2013,McQuillan2014}. This inverse correlation is also consistent with the behavior of stars in the unsaturated regimes for other activity indicators. At $\rm Ro\lesssim 0.4$, the picture is less clear. Here, the variability amplitudes no longer follow the simple inverse correlation seen in the $\rm Ro\gtrsim 0.4$ stars. This could be an indication that the variability amplitudes of these stars are beginning to saturate although we note that the transition between the saturated and unsaturated regimes generally occurs at a lower Rossby number than $\rm Ro=0.4$ in other activity indicators. Additionally, the variability amplitudes of the $\rm Ro\lesssim 0.4$ stars do not show a flat plateau, as one might expect if they were part of the saturated regime. Instead, they form a dip, or a V shape, centered at $\rm Ro\sim 0.3$. This feature can also be seen in the variability amplitude vs period plots of \citet{McQuillan2014}. It is unclear what the physical origin of this dip is although the work of \citet{Reinhold2019} may offer an explanation. These authors showed that there is valley of stars with lower variability amplitudes in their period-color diagram (their figure 9) that extends from $\sim 15$ days at B-V$\sim$0.9 mag to $\sim 25$ days at B-V$\sim$1.5 mag. We also see a similar valley of low variability in our period-mass diagram although we have not colour coded our points in fig. \ref{fig:PerMass} by variability because it makes the figure overly cluttered. This valley of low variability is the same feature as the dip at a Rossby number of $\sim$0.3 seen in fig. \ref{fig:ActRoss}. \citet{Reinhold2019} suggest that this valley is caused by bright faculae partially cancelling out dark spots resulting in a reduction in the overall variability amplitude. The difficulty in identifying the transition between the saturated and unsaturated regimes in fig. \ref{fig:ActRoss} is further compounded by the fact that our sample does not extend down much below $\rm Ro\lesssim 0.2$. Do the variability amplitudes of $\rm Ro\lesssim 0.2$ stars continue to increase with decreasing Rossby number or do they remain constant at around $R_{\rm per}\sim 10^4$ppm? The former could indicate that the dip at $\rm Ro\sim 0.3$ is just a kink in the unsaturated regime with the transition to the saturated regime occurring at a smaller Rossby number. The latter could indicate that the transition to saturation does indeed begin at a higher Rossby number, $\rm Ro\sim 0.4$, than for other activity indicators. Lastly, it may also be the case that the paradigm of saturated and unsaturated regimes does not easily apply to the variability amplitude because it is a less direct proxy of magnetic activity than many other indicators.

Another interesting feature of fig. \ref{fig:ActRoss} is the relatively sharp upper edge present in the data and the comparatively large amount of scatter in the data points below the main bulk of the variability-Rossby number relation. We propose that this is caused by variations in the angle between the observer's line of sight and the rotation axis of the star. On rotational time-scales, stars viewed equator-on will have larger variability amplitudes than those viewed pole-on \citep{Shaprio2016}. As such, two stars with the same Rossby number, and therefore same overall activity level, can have different variability amplitudes if they have different inclination angles. We therefore suggest that the upper edge in fig. \ref{fig:ActRoss} consists of stars that are viewed almost equator-on, i.e. the inclination that maximises the variability for a given level of activity/spot coverage, while stars viewed more pole-on scatter down from this maximal value resulting in a less well defined lower `edge' to the relationship. However, it is not currently possible to verify this interpretation since inclinations are not available for this sample of stars.

Lastly, we note that there is a still a residual metallicity dependence present in fig. \ref{fig:ActRoss}. At a given Rossby number, more metal-rich stars appear to have a larger variability amplitude at a given Rossby number (visible as a vertical color gradient in the figure), although there is a lot of scatter. If the influence of metallicity on photometric variability could be completely attributed to just structural effects, then we would not expect to see any metallicity dependence in fig. \ref{fig:ActRoss}. We suggest two possible explanations. The first is related to the way we have calculated our turnover times. Since convective turnover times are not observable, their estimates are model dependent \citep[see e.g.][and also appendix \ref{sec:Appendix} for further discussion]{Charbonnel2017}. It is possible that the method we used to derive a turnover timescale from the structure models involves several assumptions which could result in a slight metallicity gradient in fig. \ref{fig:ActRoss}. The second possible explanation is that metallicity may have additional effects on top of its impact on the convective turnover time. For example, the strength of Fraunhofer lines are affected by metallicity which can affect a star's variability amplitude \citep{Witzke2018}. To further investigate this, we divided the periodic sample into a number of sub-samples of approximately constant Rossby number. When examining plots of variability against metallicity for these sub-samples (not shown), we found that there was a positive correlation but that there was a lot of scatter in the plots and the correlation was always weak. This indicates that the main way that metallicity influences photometric variability is through its impact on stellar structure, with other effects having a secondary role.

\begin{figure}
	\begin{center}
	\includegraphics[trim=0cm 1.2cm 0cm 0.6cm,width=\columnwidth]{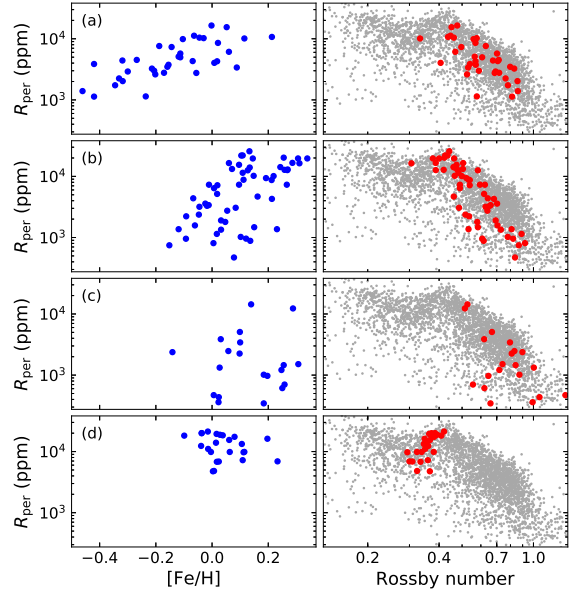}
	\end{center}
	\caption{Variability amplitude vs metallicity (left) and Rossby number (right) for 4 approximately constant period-mass bins. Each bin and the corresponding sub-panel in fig. \ref{fig:Grid} are labelled (a)-(d). The full sample is also shown in grey in the right hand plots.}
	\label{fig:IndBins}
\end{figure}

\begin{figure*}
	\begin{center}
	\includegraphics[trim=0.8cm 0.8cm 0.8cm 0cm,width=\textwidth]{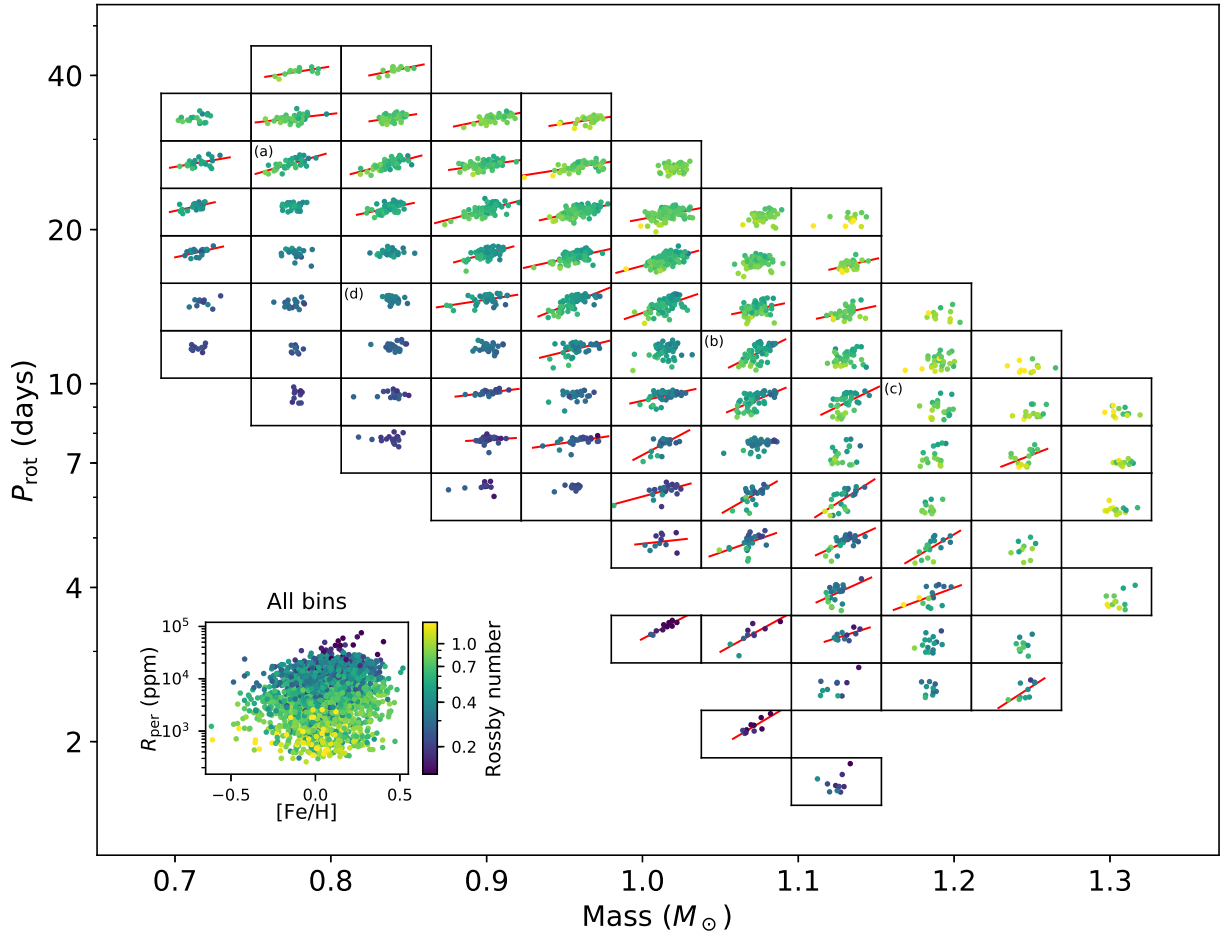}
	\end{center}
	\caption{Our sample of stars in period-mass space. Each sub-panel within the main figure represents a bin in period-mass space that contains at least 10 stars. These are the bins shown in red in fig. \ref{fig:PerMass}. Within each sub-panel, we show the photometric variability amplitude against the metallicity of the stars in that bin. The inset in the bottom left shows the variability amplitude against metallicity of all the bins, i.e. all the sub-panels stacked on top of each other. The axes of each sub-panel have the same range as the inset. All data points are coloured by Rossby number. Best fit lines are shown in red for select bins (see text). The four bins from fig. \ref{fig:IndBins} are labelled (a)-(d).}
	\label{fig:Grid}
\end{figure*}

\subsection{Trends with metallicity}
\label{subsec:MetTrends}
In order to study how the variability amplitude depends on metallicity, we first need to remove the impact of other variables such as mass or rotation. To do this, we divide the periodic sample into bins in period-mass space of size $\Delta \log P_{\rm rot} = 0.093\bf {\rm dex}$ and $\Delta M_\star = 0.058M_\odot$. These bins are indicated in figure \ref{fig:PerMass} by dashed gray lines. All the stars within a given bin will have approximately the same mass and rotation period. Any variations in the variability amplitude should therefore be due to variations in the metallicity.

In fig. \ref{fig:IndBins}, we show the variability amplitude against metallicity and rossby number for four example bins. Additionally, in fig. \ref{fig:Grid}, we plot the variability amplitude against metallicity for every bin with at least 10 stars in it (these bins are indicated in red in figure \ref{fig:PerMass}). Each of these $R_{\rm per}$ vs $\rm [Fe/H]$ plots is shown in the location of their bin in period-mass space, i.e. the location of the red bins in fig. \ref{fig:PerMass}. The inset in the bottom left of fig. \ref{fig:Grid} shows $R_{\rm per}$ vs $\rm [Fe/H]$ for all bins. The axes for each of the individual $R_{\rm per}$ vs $\rm [Fe/H]$ plots in fig. \ref{fig:Grid} have the same limits as the axes of the inset. All stars are colour coded by their Rossby number. Although we have binned our sample in period and mass space, each bin still has a small spread in period and mass. Therefore, we also visually checked $R_{\rm per}$ vs $M_\star$ and $R_{\rm per}$ vs $P_{\rm rot}$ plots for every bin to make sure that the activity trends we discuss in this section are not driven by this spread.

\begin{figure}
	\begin{center}
	\includegraphics[trim=0cm 0.8cm 0.8cm 0cm,width=\columnwidth]{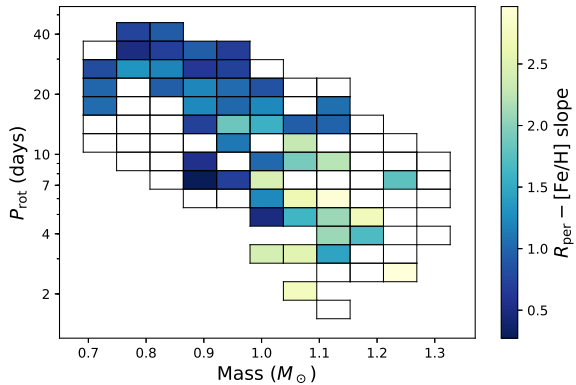}
	\end{center}
	\caption{Colour map showing the slopes, $m$, of the $R_{\rm per} - [{\rm Fe/H}]$ fit lines shown in fig. \ref{fig:Grid}. Bins that do not have a strong correlation between $R_{\rm per}$ and $\rm [Fe/H]$ are left empty (see text).}
	\label{fig:Slope}
\end{figure}

Figure \ref{fig:Grid} shows that variability amplitude and metallicity are generally positively correlated in any given bin although there can be significant scatter. Figures \ref{fig:IndBins}(a) and \ref{fig:IndBins}(b) show this correlation in more detail for two example bins ($M_\star\sim 0.78M_\odot$, $P_{\rm rot} \sim$27 days and $M_\star \sim 1.07M_\odot$, $P_{\rm rot}\sim$11 days respectively). The positive correlation between variability amplitude and metallicity matches the theoretical expectation we outlined in the introduction that, all else being equal, more metal-rich stars have smaller Rossby numbers and should therefore be more magnetically active. Indeed, the right hand panels of figs. \ref{fig:IndBins}(a) and \ref{fig:IndBins}(b) show that the most metal rich stars in those bins have the smallest Rossby numbers. A Rossby number gradient can also be seen across each of the $R_{\rm per}$ vs $\rm [Fe/H]$ plots in fig. \ref{fig:Grid} from the colour gradient present. It is worth reiterating that the majority of the spread in Rossby numbers, and therefore variability amplitudes, in each bin is due to the spread in metallicity rather than the spread in mass or period.

Although the variability amplitude is generally positively correlated with metallicity, there are regions of parameter space where this is not the case. The first region is stars with large Rossby numbers, Ro$\gtrsim$ 0.9. Figure \ref{fig:Grid} shows that many of the $R_{\rm per}$ vs $\rm [Fe/H]$ plots in this Rossby number regime, i.e. those at high mass and long rotation period, have a lot of scatter and show no clear trend. Looking at fig. \ref{fig:ActRoss}, one can see that it is more sparsely populated at Ro$\gtrsim$ 0.9. Additionally, the stars in this regime have more scatter than those at smaller Rossby numbers, making it harder to identify any trends. The large amount of scatter, both in $R_{\rm per}$ vs [Fe/H] space and $R_{\rm per}$ vs Rossby number space, can also clearly be seen in fig. \ref{fig:IndBins}(c) which is a typical bin in the high Rossby number regime ($M_\star \sim$1.18$M_\odot$, $P_{\rm rot}\sim$9 days). It is unclear whether the large amount of scatter is intrinsic or whether the trends would be improved if this high Rossby number regime could be better sampled. The second regime where $R_{\rm per}$ and $\rm [Fe/H]$ are not clearly positively correlated is stars at low Rossby numbers, Ro$\lesssim$ 0.4, i.e. those at low mass and rapid rotation in figure \ref{fig:Grid}. The reason for this is the dip seen in fig. 2 at Ro$\sim$0.3. Since the stars in this Rossby number regime do not follow the overall trend of increasing variability amplitude with decreasing Rossby number, the positive correlation between $R_{\rm per}$ and $\rm [Fe/H]$ is also not seen for these stars. Figure \ref{fig:IndBins}(d), which is a bin in this low Rossby number regime ($M_\star \sim$0.84$M_\odot$, $P_{\rm rot}\sim$14 days), demonstrates this behaviour clearly. Although we do not see a positive correlation between metallicity and variability amplitude for Ro$\lesssim$ 0.4 stars, it may be the case that a positive correlation would exist in this region of parameter space if a different activity proxy were used (see section \ref{subsec:OtherActInd}).

As well as showing the correlation between variability amplitude and metallicity itself, fig. \ref{fig:Grid} also suggests that there may be a mass dependence in this relationship. Specifically, there are hints that the slope of the $R_{\rm per}$ vs  $\rm [Fe/H]$ relationship is steeper for higher mass stars. We can see this trend when we compare figs. \ref{fig:IndBins}(a) and \ref{fig:IndBins}(b). These bins were chosen to have a comparable range of Rossby numbers but different masses. Bin (b) ($\sim$1.07$M_\odot$) clearly has a steeper $R_{\rm per}$ vs [Fe/H] relationship than bin (a) ($\sim$0.78$M_\odot$). We also calculate best fit lines, of the form $\log R_{\rm per} = m[{\rm Fe/H}]+c$ for each of the bins in fig. \ref{fig:Grid}. As already noted, some regions of parameter space do not show strong correlations between variability and metallicity and so we only plot the best fit lines for the bins where the variability vs metallicity relationship has a Spearman's correlation coefficient greater than 0.3. For clarity, we show the slopes, $m$, of these fits as a colourmap in fig. \ref{fig:Slope}. Although there is a significant amount of scatter present, there is a rough trend of larger gradients in the higher mass bins. Previous theoretical work using stellar structure models has demonstrated that the convective turnover time, and hence Rossby number, is more sensitive to changes to metallicity in higher mass stars \citep{AmardMatt2020} which may explain why variability amplitude appears to be have a stronger dependence on metallicity in higher mass stars.

\section{Rotation period detection}
\label{sec:RotRecovery}
Techniques like Lomb-Scargle periodograms or autocorrelation methods are often used to recover rotation periods from photometric light curves. However, the variability in the light curve must exceed some threshold level for these techniques to work. In section \ref{sec:Trends}, we showed that variability is generally correlated with metallicity. As such, we should expect that metallicity will also affect how easily rotation periods can be detected. The metallicitiy histogram in fig. \ref{fig:Hists} already gives us a hint of this effect. It shows that there are a lot more stars for which periods could not be detected at low metallicities, [Fe/H]$\lesssim 0$.

\begin{figure}
	\begin{center}
	\includegraphics[trim=0cm 0.8cm 0.8cm 0cm,width=\columnwidth]{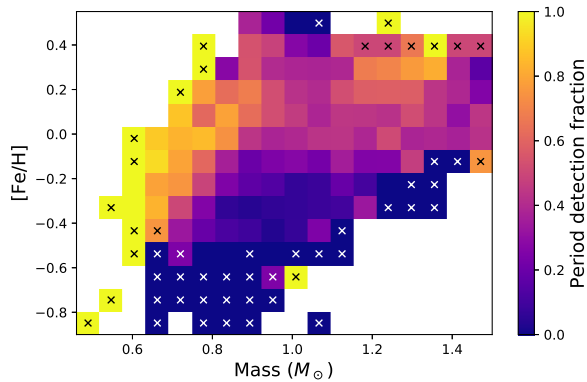}
	\end{center}
	\caption{The fraction of stars for which a rotation period is detected in mass and metallicity bins. Bins containing 5 or fewer stars are marked with a cross (the colour of the crosses only change to improve visibility and have no additional meaning).}
	\label{fig:PeriodRecovery}
\end{figure}

To further investigate this, we combine the periodic and non-periodic samples together to look at the fraction of stars for which a period is detected as a function of mass and metallicity. Figure \ref{fig:PeriodRecovery} shows the period detection fraction for our sample in mass and metallicity bins. Bins that contained 5 or fewer stars are marked with crosses and may deviate from the overall trends shown by the other bins due to low number statistics. An obvious feature of fig. \ref{fig:PeriodRecovery} is the complete lack of either periodic or non-periodic stars in the bottom right of the plot, i.e. at low metallicity and high mass. This lack of stars can be attributed to the fact that \citet{McQuillan2014} eliminated any star hotter than 6500 K from their sample (and recalling that at fixed effective temperature, a more metal-rich star must be higher mass). In the regions of metallicity-mass parameter space that are populated, the period detection fraction is generally highest for lower mass stars and more metal-rich stars (although there is a hint that $M_\star \sim 1M_{\odot}$ and [Fe/H]$\sim$0.5 stars may break this overall trend). This reflects the fact that lower mass stars are generally more active than higher mass stars \citep[e.g.][]{Wright2018} and that metal-rich stars are generally more active than metal-poor stars (as demonstrated in section \ref{subsec:MetTrends}). Therefore, the rotation periods of low-mass, metal-rich stars are easier to measure because of their larger variability amplitudes. Indeed, this metallicity dependent bias in the context of detecting rotation periods has previously been suspected by \citet{Claytor2020}. Additionally, the mass dependence could explain why \citet{McQuillan2014} found that their period detection fraction was higher for cooler stars. As can be seen in fig. \ref{fig:PeriodRecovery}, rotation periods can be reliably detected in low-mass stars (corresponding roughly to cooler stars) down to lower metallicity. However, this comparison is complicated by the fact that we focus on mass while \citet{McQuillan2014} focus on effective temperature.

An interesting feature of this sample is the fact that the lower limit on variability amplitudes that periods could be measured for is dependent on effective temperature. For the hottest stars in their sample ($T_{\rm eff} \gtrsim 5400$K), \citet{McQuillan2014} could measure periods for stars with variability amplitudes down to their detection threshold of $\sim$300ppm. However, for cooler stars ($T_{\rm eff} \lesssim 5400$K), they could only measure periods for stars with variability amplitudes down to $\sim$2000ppm which is significantly higher than their detection threshold (see their fig. 3). These authors suggested that the lack of cooler stars with variability amplitudes below 2000ppm is related to inclination effects. While inclination effects likely play some role, we offer an additional suggestion. Taken at face value, our work suggests that it should have been possible for periods to be measured in $T_{\rm eff} \lesssim 5400$K stars with variability amplitudes smaller than $\sim$2000ppm provided they were sufficiently metal-poor. We therefore propose that the reason for the lack of $T_{\rm eff} \lesssim 5400$K, $R_{\rm per} \lesssim$ 2000ppm stars in the periodic sample is that the $T_{\rm eff} \lesssim 5400$K stars do not extend down to low enough metallicity. We can look at the non-periodic sample to see if its properties are consistent with this idea. Indeed, the vast majority of the non-periodic sample have [Fe/H]$>$-0.5 which can be seen in fig. \ref{fig:Hists}. The linear fits for the coolest, or least massive stars, in fig. \ref{fig:Grid} suggest that they would require metallicities lower than -0.5  to have variability amplitudes significantly lower than 2000ppm. This is can be seen especially clearly in fig. \ref{fig:IndBins}(a). However, we again note that this suggestion is muddied by the fact that we use mass while \citet{McQuillan2014} use effective temperature.

Lastly, it is worth comparing fig. \ref{fig:PeriodRecovery} with the work conducted by \citet{Witzke2020}. These authors also investigated the effect of metallicity on period detection but restricted their analysis to a sample of solar-like Kepler stars. They found that the period detection fraction is lowest in stars with approximately solar metallicity with the detection fraction increasing for more metal-poor and metal-rich stars. Looking at the solar mass bins in fig. \ref{fig:PeriodRecovery}, the period detection fraction exhibits a maximum at around solar metallicity which seems to be incompatible with the result of \citet{Witzke2020}. However, there are a few reasons that could explain this difference. The first is that the sample of stars studied by \citet{Witzke2020} was selected on the basis of effective temperature (5600K-5900K) whereas we use mass. As already noted, this complicates the comparison between our two works since metallicity and effective temperature are not independent for a given mass. Additionally, \citet{Witzke2020} made a $R_{\rm per}<0.18\%$ cut to their sample and only included stars with periods between 24-30 days in their periodic sample whereas we do not make cuts in variability amplitude or period for fig. \ref{fig:PeriodRecovery}. The different sample selection criteria likely explain the difference between the period detection trend found by \citet{Witzke2020} and the one in our solar mass bins in fig. \ref{fig:PeriodRecovery}. 

\section{Further discussion}
\label{sec:Disc}

\begin{figure}
	\begin{center}
	\includegraphics[trim=0cm 0.8cm 0cm 0cm,width=\columnwidth]{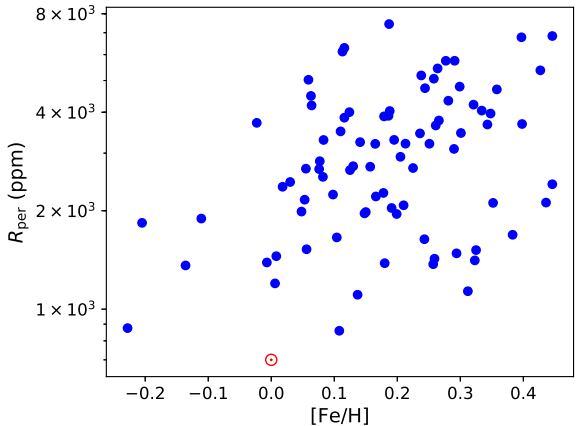}
	\end{center}
	\caption{Variability amplitude vs metallicity for all the stars in our sample with solar-like properties ($0.95M_\odot < M_\star < 1.05M_\odot$, $24\ {\rm days} < P_{\rm rot} < 30\ {\rm days}$). The Sun is shown with a red solar symbol.}
	\label{fig:SolarBin}
\end{figure}

\subsection{The solar variability in a stellar context}
\label{subsec:LowSolar}
It has previously been noted that the Sun has a very low variability amplitude when compared to stars with similar masses (or effective temperatures) and rotation periods \citep{Reinhold2020}. To investigate this, we can look at the Sun's variability in the context of our sample. In fig. \ref{fig:SolarBin}, we show variability amplitude against metallicity for all the stars in our sample with solar-like properties ($0.95M_\odot < M_\star < 1.05M_\odot$, $24\ {\rm days} < P_{\rm rot} < 30\ {\rm days}$). Note that this is a larger bin size than that used for fig \ref{fig:Grid}. As already demonstrated in section \ref{subsec:MetTrends}, more metal-rich stars generally have larger variability amplitudes. For the solar variability amplitude, we adopt a median value of 700ppm from \citet{Reinhold2020}. Similarly to \citet{Reinhold2020}, we find that the solar variability is very low when compared to other solar-like stars. Figure \ref{fig:SolarBin} shows that the Sun falls within the trend set by the rest of the stars in its bin but lies at the lower end of the scatter. However, our work suggests that solar-like stars that are sufficiently metal-poor should have variability amplitudes that are smaller than the solar variability amplitude. If such metal-poor stars can be found and their variability measured, the Sun's variability amplitude may turn out to be more typical.

\subsection{Implications for rotation evolution}
\label{subsec:RotEvo}
It is well known that the rate at which a star loses angular momentum depends on its magnetic activity, specifically its mass-loss rate and magnetic field strength \citep{Matt2012,Vidotto2014,Garraffo2015,Reville2015,Pantolmos2017,Finley2018}. A number of previous rotation evolution models have made the assumption that these forms of activity can be parameterised in terms of just the Rossby number \citep{vanSaders2013,Matt2015}. This assumption implies that mass-loss rates and field strengths, or some combination of the two, should be larger for more metal-rich stars. As such, more metal-rich stars spin down faster in these models \citep{AmardMatt2020}. Our study takes a step towards validating this assumption by demonstrating that more metal-rich stars generally have larger variability amplitudes. Since variability amplitude, mass-loss rate and magnetic field strengths are all forms of magnetic activity, it seems likely that mass-loss rates and field strengths should also be stronger in more metal-rich stars, i.e. the behaviour assumed in these rotation evolution models. Additionally, recent work studying stars in the Kepler field has shown that more metal-rich stars appear to spin slower on average \citep{Amard2020}. Although, we cannot currently directly test how mass-loss rates or magnetic field strengths depend on metallicity, this observational evidence from \citet{Amard2020}, together with the results we present in this work, strongly suggests that spin-down does depend on metallicity in the manner implied by these rotation evolution models.

\subsection{Other activity proxies}
\label{subsec:OtherActInd}
Given that photometric variability seems to correlate with overall activity levels \citep{Radick1998,Karoff2016,Salabert2016}, one might expect that other proxies of magnetic activity should also grow stronger with increasing metallicity. Indeed, repeating the type of study we have presented using other activity proxies, e.g. X-ray emission, or even direct magnetic field measurements, would be an interesting exercise since they are more direct tracers of the underlying dynamo. It would also be a direct confirmation of the assumptions made by some rotational evolution modelling (see section \ref{subsec:RotEvo}). Previous works have shown that other proxies of activity monotonically increase with decreasing Rossby number in the unsaturated regime \citep{Saar1999,Reiners2009,Stelzer2016,Wright2018,See2019ZB}. This is in contrast to the variability amplitude which has a dip at $\rm Ro \sim 0.3$ as shown in fig. \ref{fig:ActRoss}. As such, the activity vs metallicity trends, i.e. plots corresponding to those shown in fig. \ref{fig:Grid} but for these more direct tracers, may be even more clear than the ones we have shown for the variability amplitude. Additionally, other activity proxies will not have such a strong dependence on inclination as the variability amplitude. Again, this should increase the strength of any activity vs metallicity trends. However, it may be some time before such studies are feasible. The work we have presented here is only possible because of the large number of well characterised stars for which variability amplitudes are available \citep{McQuillan2014} and there is currently no comparable data set for other activity indicators. 

\subsection{The saturated regime}
\label{subsec:SatRegime}
One region of parameter space that our study has not been able to probe is the saturated regime. Previous studies show that the activity indicators of stars with Rossby numbers smaller than some critical value saturate to a constant level, i.e. activity is independent of Rossby number \citep{Reiners2009,Stelzer2016,Wright2018,See2019ZB}. Since our results suggest that the impact of metallicity on activity can be explained by its impact on the turnover time, and therefore the Rossby number, we would expect that metallicity should not have any effect on activity in the saturated regime. Assuming that the critical Rossby number at which stars transition from the saturated to unsaturated regimes is fixed, this leads to an interesting implication. For a fixed mass, the critical rotation period should be longer for more metal-rich stars to compensate for their longer convective turnover times. Such an effect has previously been discussed in a rotation evolution context \citep{AmardMatt2020}.

\section{Summary}
\label{sec:Summary}
In this work, we have studied the relationship between magnetic activity and metallicity, using photometric variability as an activity proxy. This was done with a sample of over 3000 stars for which we had estimates of metallicities, rotation periods, variability amplitudes, masses and convective turnover times. The first three parameters were taken from the literature \citep{McQuillan2014,Luo2019} while the latter two were calculated using a grid of stellar structure models \citep{Amard2019}. 

We first demonstrated that, similarly to other activity proxies, photometric variability amplitudes can be parameterised by the Rossby number. In general, stars with smaller Rossby numbers have larger variability amplitudes. However, there is some additional structure in the variability vs Rossby number relationship that is not seen in other activity-rotation relations. We then showed that the photometric variability amplitude is generally positively correlated with metallicity for stars at approximately fixed mass and rotation period. This can be understood in terms of the impact that metallicity has on the internal structure of low-mass stars. More metal-rich stars have deeper convection zones and longer convective turnover times. Therefore, their Rossby numbers are smaller and they have stronger dynamos which results in stronger activity levels. 

These results represent a significant step forwards in understanding how metallicity affects the generation of magnetic activity in low-mass stars. They also have implications for a number of areas of stellar astrophysics. In particular, we analysed the impact of stellar metallicity on how easy it is to detect rotation periods using stellar light curves. We showed that rotation periods are more easily measured for lower mass and more metal-rich stars because variability amplitudes are largest in these stars.

\section*{Acknowledgments}
We thank an anonymous referee for valuable comments that have significantly improved the quality of this manuscript. All authors acknowledge funding from the European Research Council (ERC) under the European Union's Horizon 2020 research and innovation programme (grant agreement No 682393 AWESoMeStars). 

\emph{Software:} \texttt{astropy} \citep{astropy2013,astropy2018}, \texttt{matplotlib} \citep{Hunter2007}, \texttt{numpy} \citep{Harris2020}, \texttt{scipy} \citep{Virtanen2020}, \texttt{TOPCAT} \citep{Taylor2005}

\newpage

\appendix
\section{Estimating convective turnover times}
\label{sec:Appendix}
Estimating convective turnover times (and the associated Rossby numbers) is important for our work. Given the difficulty in determining turnover times, it will be worth comparing our estimates to some of the more commonly used ones in the literature and also discussing some of their drawbacks. Specifically, we will compare to the methods given by \citet{Noyes1984} and \citet{Cranmer2011}. These works give analytic expressions that specify the turnover time as a function of B-V colour and effective temperature respectively. To compare our turnover times to these analytic expressions, we use a set of structure models from the grid of \citet{Amard2019} over the mass range 0.6$M_\odot$ - 1.4$M_\odot$ (which covers the majority of our periodic sample), for $\rm [Fe/H]=\{-0.3,0.0,+0.3\}$ and at an age of 1 Gyr. For each model in this parameter space, we output the turnover time as described in section \ref{sec:Sample} as well as the B-V colour and the effective temperature. We then calculate turnover times using the expressions given by \citet{Noyes1984} and \citet{Cranmer2011} from the output B-V colours and effective temperatures.

\begin{figure}[h!]
	\begin{center}
	\includegraphics[trim=0cm 1cm 0cm 0cm,width=\textwidth]{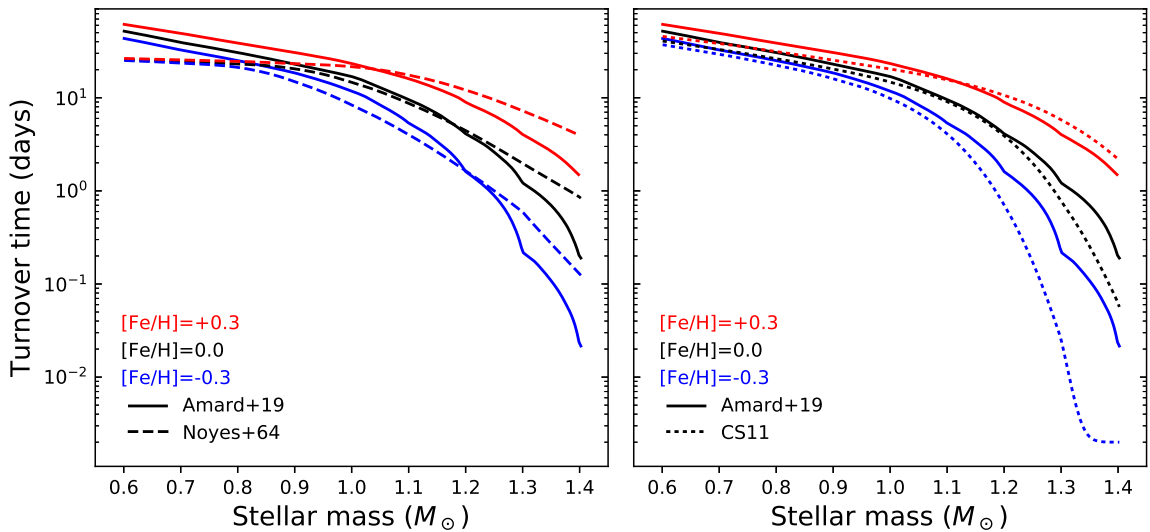}
	\end{center}
	\caption{Comparison of the convective turnover times used in this work (solid lines), derived from the structure models of \citet{Amard2019}, with the methods of \citet[][left, dash lines]{Noyes1984} and \citet[][right, dotted lines]{Cranmer2011} as a function of stellar mass. In each case, the models are compared for three metallicities, $\rm [Fe/H]=\{-0.3,0.0,+0.3\}$, in blue, black and red respectively.   }
	\label{fig:tauComp}
\end{figure}

In fig. \ref{fig:tauComp}, we compare the turnover times from the models of \citet[][solid lines]{Amard2019} to those estimated using the formulas from \citet[][dashed lines]{Noyes1984} and \citet[dotted lines]{Cranmer2011}. The turnover time estimates monotonically decrease as a function of stellar mass at fixed metallicity and increase for increasing metalllicity at fixed mass for all three methods. The agreement between \citet{Amard2019} and the other two methods is best between $\sim 1{\rm M}_\odot$ and $\sim 1.2{\rm M}_\odot$. Below these masses, the agreement with the \citet{Cranmer2011} method is still relatively good but the \citet{Noyes1984} estimates are systematically smaller than those from \citet{Amard2019}. However, the \citet{Noyes1984} method is relatively unconstrained at lower masses. Specifically, these authors note that their formula is poorly defined for $B-V>1$ mag, corresponding approximately to $\rm M_\star \lesssim 0.8 M_\odot$ at solar metallicity, due to a lack of data in this region. At masses above $\rm \sim 1.2 M_\odot$, the agreement is also less good between the three estimates. The \citet{Noyes1984} estimates are larger than the \citet{Amard2019} estimates for all three metallicities while the \citet{Cranmer2011} estimates are smaller than the \citet{Amard2019} estimates for the $\rm [Fe/H] =$ 0.0 and -0.3 cases. It is worth noting that the convective envelope becomes extremely thin at the highest masses. As such, the convective zone depth becomes very sensitive to differences in the input physics of the models that the different methods are based on. This likely explains the discrepancies seen between the turnover time estimates at high mass. 

Another factor to consider is the use of mixing-length theory in the models of \citet{Amard2019} (and also the models of \citet{Gilman1980} and \citet{Gunn1998} which underpin the \citet{Noyes1984} and \citet{Cranmer2011} prescriptions respectively). 
Although mixing length theory can reproduce many observations, it is a very simple description of convection and does not account for dependencies that convection should have on global stellar properties \citep{Trampedach2014}. Additionally, it is common to have a fixed mixing-length parameter, $\alpha$, that is calibrated to a specific star (usually the Sun) when using mixing-length theory. However, using well constrained asteroseismic targets, \citet{JoyceChaboyer2018} demonstrated that the value of $\alpha$ is affected by the choice of calibration star. Similarly, \cite{Viani2018} used the mixing-length values obtained from asteroseismology to find empirical relationships with surface gravity and metallicity. 
Finally, $\alpha$ may very well vary with depth \citep{IrelandBrowning2018}. Most of these works point out the relative importance of metallicity and surface gravity for the value of $\alpha$, and show a more efficient convection with higher metallicity (higher $\alpha$) thus changing the convective turnover timescale in a complex way \citep{Valle2019}.

Our comparison of several common methods of estimating convective turnover times show that they all give slightly different turnover time estimates but follow the same overall trends as a function of mass and metallicity. Using a different method would therefore move the datapoints in fig. \ref{fig:ActRoss} around slightly but it would not change the overall qualitative shape of the plot. As such, the main conclusions of this paper are robust to our choice of method for estimating turnover times, at least among these commonly used prescriptions. However, it is worth keeping in mind that these methods of estimating turnover times are still relatively simple. Exploring the effect of more sophisticated treatments of convection is outside the scope of this work but represents an interesting avenue of exploration for future work.

\bibliography{bibfile}{}
\bibliographystyle{aasjournal}

\end{document}